%% file: charon.tex
\documentclass[conference]{IEEEtran}
\IEEEoverridecommandlockouts
\usepackage{fancyhdr} 
\usepackage{cite}
\usepackage{amsmath,amssymb,amsfonts}
\usepackage{algorithmic}
\usepackage{graphicx}
\usepackage{textcomp}
\usepackage{xcolor}
\usepackage{mathtools}
\usepackage{tikz}
\usepackage{enumitem}
\usepackage[hyphens]{url}                        
\usepackage[hidelinks]{hyperref}                 
\usepackage[colorinlistoftodos,prependcaption,textsize=small]{todonotes} 
\usepackage{booktabs}				
\usepackage{multirow}
\usepackage{makecell} 				  
\usepackage{atbegshi,picture} 
\makeatletter\let\MYcaption\@makecaption\makeatother
\usepackage[font=footnotesize]{subcaption}
\makeatletter\let\@makecaption\MYcaption\makeatother

\newcommand*{\myfont}{\fontfamily{lmss}\selectfont}

\AtBeginShipout{\AtBeginShipoutUpperLeft{%
  \put(\dimexpr\paperwidth-5cm\relax,-.5cm){\makebox[0pt][r]{{\small \myfont 2021 1st Joint International Workshop on Network Programmability and Automation}}}%
}}

\def\BibTeX{{\rm B\kern-.05em{\sc i\kern-.025em b}\kern-.08em
    T\kern-.1667em\lower.7ex\hbox{E}\kern-.125emX}}

\fancypagestyle{specialfooter}{%
    
    \fancyfoot[L]{{\small \myfont 978-3-903176-36-2 \textcopyright 2021 IFIP}}
  }

\begin{document}

\newcommand{\Albatross}{Charon}
\newcommand{\figwidth}{0.90\linewidth}
\newcommand{\red}[1]{{\color{red}{#1}}}
\newcommand{\E}{\mathbb{E}}
\newcommand{\ie}{\textit{i.e.}\ }
\newcommand{\eg}{\textit{e.g.}\ }
\newcommand{\wrt}{\textit{w.r.t.}\ }
\newcommand{\etc}{etc.\ }
\newcommand{\publicationtype}{paper}
\newcommand{\Publicationtype}{Paper}
\newcommand{\var}[1]{{\ttfamily#1}}
\newcommand*\circled[1]{\tikz[baseline=(char.base)]{
		\node[shape=circle,draw,inner sep=1pt] (char) {#1};}}
\renewcommand{\figurename}{Figure}

\title{\Albatross: Load-Aware Load-Balancing in P4}

\author{Carmine Rizzi, Zhiyuan Yao$^{\ast}$, Yoann Desmouceaux, Mark Townsley, Thomas Clausen
\thanks{C. Rizzi, Z. Yao and T. Clausen are with \'Ecole Polytechnique, 91128 Palaiseau, France; emails \{carmine.rizzi,zhiyuan.yao,thomas.clausen\}@polytechnique.edu.}
\thanks{C. Rizzi, Z. Yao, Y. Desmouceaux and M. Townsley are with Cisco Systems Paris Innovation and Research Laboratory (PIRL), 92782 Issy-les-Moulineaux, France; emails \{crizzi, yzhiyuan, ydesmouc,townsley\}@cisco.com.}
\thanks{$\ast$Z. Yao is the corresponding author.}
}

\maketitle

\begin{abstract}
\input{content/0-abstract.tex}
\end{abstract}


\thispagestyle{specialfooter}
\input{content/1-intro.tex}

\input{content/2-overview.tex}

\input{content/3-design.tex}

\input{content/4-implement.tex}

\input{content/5-evaluation.tex}

\input{content/6-conclusion.tex}

\bibliographystyle{IEEEtran}
\bibliography{reference.bib}

\end{document}

%% file: content/0-abstract.tex
Load-Balancers play an important role in data centers as they distribute network flows across application servers and guarantee per-connection consistency.
It is hard however to make fair load balancing decisions so that all resources are efficiently occupied yet not overloaded.
Tracking connection states allows load balancers to infer server load states and make informed decisions, but at the cost of additional memory space consumption.
This makes it hard to implement on programmable hardware, which has constrained memory but offers line-rate performance.
This paper presents \Albatross, a stateless load-aware load balancer that has line-rate performance implemented in P4-NetFPGA.
\Albatross\ passively collects load states from application servers and employs the power-of-$2$-choices scheme to make load-aware load balancing decisions and improve resource utilization.
Per-connection consistency is guaranteed statelessly by encoding server ID in a covert channel.
The prototype design and implementation details are described in this paper.
Simulation results show performance gains in terms of load distribution fairness, quality of service, throughput and processing latency.

%% file: content/1-intro.tex
\section{Introduction}
\label{sec:intro}

Data centers (DCs) have seen a rising amount of connections to manage~\cite{ananta2013, facebook-dc-architecture} and users expect an elevated server responsiveness~\cite{duet}.
Due to these conditions, applications are virtualized in replicated instances in data centers to provide scalable services~\cite{dragoni2017microservices, bernstein2014containers}.
A given service provided in a data center is identified by virtual IP (VIP).
Each application instance behind the VIP is identified by direct IP (DIP). 
In this architecture, load balancers (LBs) play an important role.
They distribute requests from clients among application servers and maintains per-connection consistency (PCC)~\cite{ananta2013, maglev}.

\begin{figure}[t]
	\centering
	\includegraphics[width=\columnwidth]{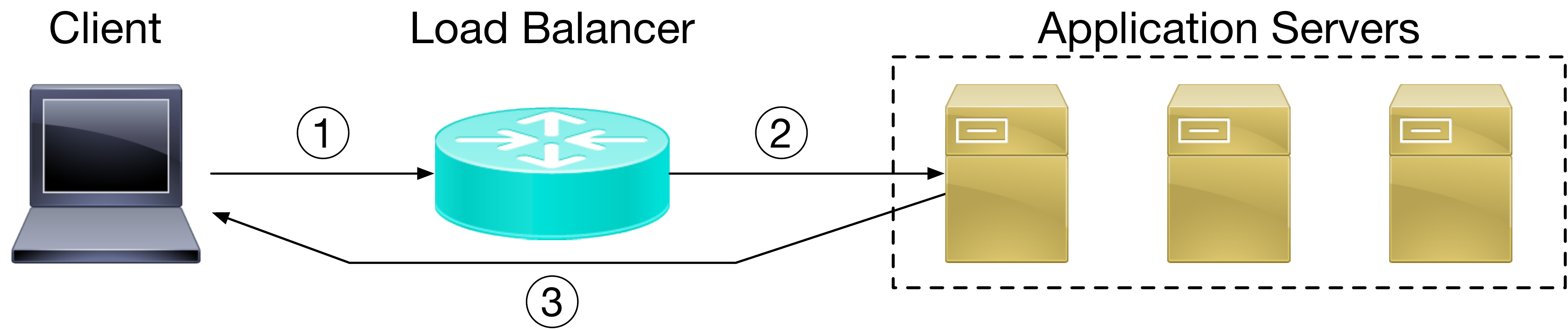}
	\caption{Network load balancer in data centers.}
	\label{fig:intro-context}
	\vskip -.1in
\end{figure}

This paper exemplifies the challenges that LBs should tackle by way of a simple heuristic load balancing mechanism, \ie Equal Cost Multi Path (ECMP).
As is depicted in figure~\ref{fig:intro-context}, on receipt of a new request (step \textcircled{1}), ECMP LBs randomly select a server among the server pool to which the request is forwarded (step \textcircled{2}), based on the hash over the $5$-tuple of the connection\footnote{The $5$-tuple corresponds to IP source, IP destination, protocol number, TCP source port and TCP destination port.}.
The replies are sent back directly to the client instead of traversing the LBs (step \textcircled{3}) in direct source return (DSR) mode.
DSR mode is first proposed in~\cite{ananta2013} so that LBs avoid handling $2$-way traffic and becoming a throughput bottleneck between servers and clients.
Though easy to implement, ECMP is agnostic to the server load states.
As ECMP randomly distribute workloads, new requests may be forwarded to overloaded servers, reducing load balancing fairness.
ECMP is also not able to guarantee PCC since server pool updates change the DIP entries in the hash table, which potentially forwards subsequent packets of established connections to different servers and breaks connections.

\subsection{Related Work}
\label{sec:intro-related}

To guarantee PCC, stateful LBs keep tracking the state of the connections~\cite{ananta2013, maglev, 6lb, spotlight2018}.
Using advanced hashing mechanism (\eg consistent hashing~\cite{maglev, 6lb}), server pool updates have little impact on the hashing table therefore the amount of disrupted connections is decreased.
However, stateful LBs require additional memory space for flow tables to store connection states.
When encountering DoS attacks, flow tables risk of being filled by malicious flows and no longer track benigh flows.
In case of LB failures, the tracked connection states are lost and all connections via the failed LB need to be re-established, which degrades quality of service (QoS).
Stateless LBs~\cite{shell2018, faild2018, beamer} use alternative mechanisms to recover the right server destinations, without keeping the flows' states.
They encode server id information in packet headers and daisy-chain two possible server candidates, to retrieve a potentially changed flow-server mappings.
\Albatross\ adopts stateless load balancing scheme~\cite{shell2018, cheetah2020, faild2018} and encapsulates the server id inside the packet. In particular, the TCP timestamp option~\cite{rfc7323} is used to transport this information. 

\begin{figure*}[t]
	\vskip -0.05in
	\centering
	\begin{subfigure}{\columnwidth}
		\centering
		\includegraphics[width=.95\columnwidth]{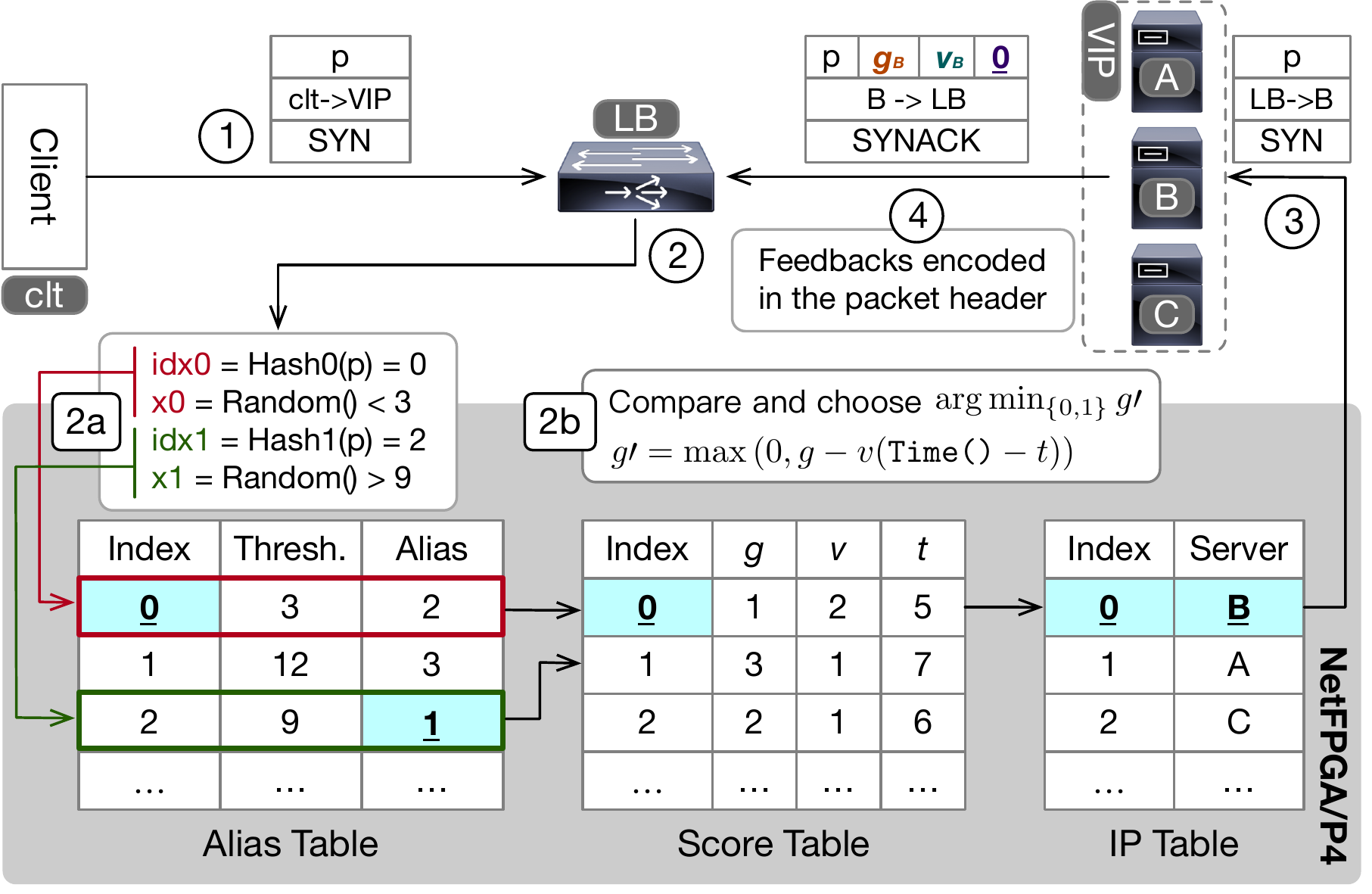}
		\vskip -0.1in
		\caption{}
		\label{fig:overview-workflow0}
	\end{subfigure}
	\hspace{.05in}
	\begin{subfigure}{\columnwidth}
		\centering
		\includegraphics[width=.95\columnwidth]{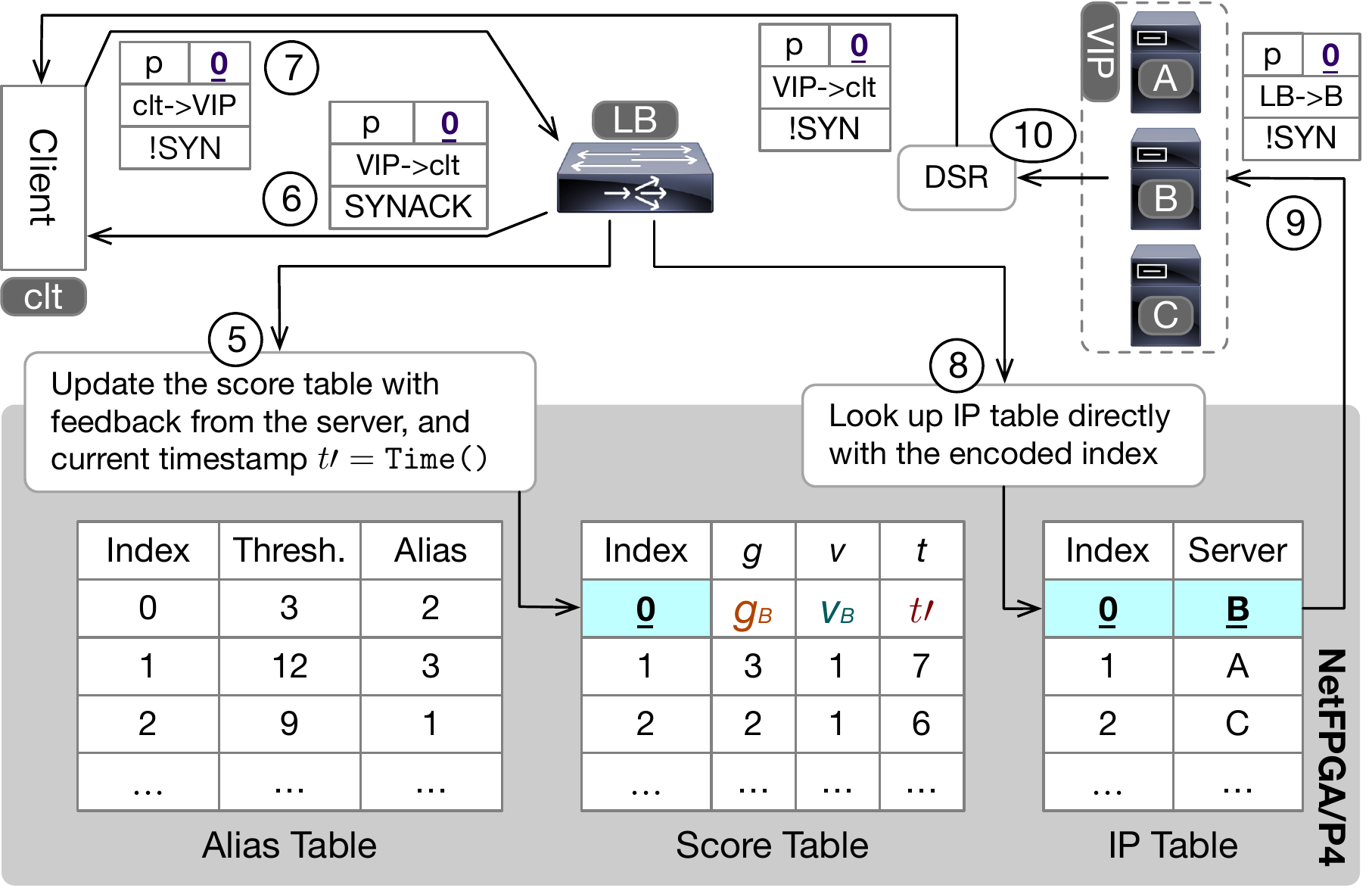}
		\vskip -0.1in
		\caption{}
		\label{fig:overview-workflow1}
	\end{subfigure}
	\caption{\Albatross\ overview.}
	\label{fig:overview-workflow}
	\vspace{-.2in}
\end{figure*}

To improve load balancing fairness, different mechanisms are proposed to evaluate server load states before making load balancing decisions.
Segment Routing (SR)~\cite{SR_rfc8402} and power-of-$2$-choice~\cite{po2choice} are used in~\cite{6lb, shell2018} to daisy chaine $2$ servers and let them decide, based on their actual load states, whether or not the new flow should be accepted.
Another approach is to periodically poll servers' instant ``available capacities''~\cite{spotlight2018}.
Ridge Regression is used in~\cite{lbas-2020} to predict server load states and compute the relative ``weight'' of each server for Weighted Costed Multi-Path (WCMP).
In~\cite{incab2018}, the servers are clustered based on their load states, where clusters with less workload are prioritized.
The servers notify the LBs about load state changes if their resource consumption surpasses pre-defined thresholds.
LVS~\cite{LVS} presents a heuristic that combines the queue lengths of active flows and provisioned server capacity to determine server load states.
Unlike prior arts, \Albatross\ passively polls and retrieves the server load when a new flow is assigned to it.
The feedback is used to predict the future server load states and make informed and fair load balancing decisions, which improves resource utilization and QoS.

To optimize performance in terms of throughput and latency, different hardware solutions are proposed.
Silkroad implements LB functions on dedicated hardware device~\cite{silkroad2017}, while other designs implement a hybrid solution combining software and hardware LBs~\cite{duet, rubik2015} to guarantee PCC.
As a hardware solution, \Albatross\ is realized on a NetFPGA board using P4-NetFPGA tool-chain~\cite{P4-NetFPGA} and achieves low jitter and delay.

\subsection{Statement of Purpose}
\label{sec:intro-statement}

This \publicationtype\ proposes \Albatross, a stateless, load-aware, hardware load balancer.
This paper targets the fore-mentioned $3$ aspects of LB performance:
\begin{itemize}
    \item \textit{Availability}: encapsulates the chosen server id in the covert channel of packet headers. Different covert channels are available (\eg \texttt{connection-id} of QUIC connections and the least significant bits of IPv6 addresses)~\cite{cheetah2020}. This paper uses the higher-bits of TCP timestamp options.
    \item \textit{Fairness}: \Albatross\ makes load balancing decisions on predicted server load states based on passive feedback from the application servers with actual load states encoded in \texttt{SYNACK} packets.
    Two factors are integrated at the same time, \ie queue lengths and processing speed.
    \item \textit{Performance}: \Albatross\ implements all functionalities on programmable hardware to boost performance and achieve low latency and high throughput.
\end{itemize}
Simulations show promising results and performance gain using \Albatross.
Physical testing also demonstrates the high throughput of the board.

\subsection{Paper Outline}
\label{sec:intro-outline}

The rest of this \publicationtype\ is organized as follows.
In section \ref{sec:overview} the overview of \Albatross\ is described. 
Section \ref{sec:design} presents the design of \Albatross in details. 
Section \ref{sec:implement} describes the implementation of \Albatross\ and section \ref{sec:evaluation} shows the evaluation results obtained.
Section \ref{sec:conclusion} concludes the \publicationtype.

%% file: content/2-overview.tex
\section{Overview}
\label{sec:overview}



\Albatross\ relies on $3$ tables and $1$ server agent to achieve stateless load-aware load balancing on NetFPGA.
$2$ tables are constructed and managed by the control plane.
The \textit{Alias Table} allows to select servers based on various weights with low computational complexity and low memory space consumption.
The \textit{IP Table} is used to map server id to actual IP address.
$1$ table, namely the \textit{Score Table}, is updated in the data plane on per-flow basis.

The workflow is exemplified in figure~\ref{fig:overview-workflow}.
When a \texttt{SYN} packet reaches the LB (step \textcircled{1}), \Albatross\ employs power-of-$2$-choices and applies $2$ hash functions to the $5$-tuple of the packet.
The $2$ hashes are then used as indexes in the ``Alias Method''~\cite{AliasMethod} (step 2a) to generate $2$ random server candidates based on their relative weights, which is explained in section~\ref{sec:design}.
Referring to the Score Table, \Albatross\ calculates and compares the load states of the $2$ candidate servers (step 2b).
The server with lower score is assigned to the new flow. 
In the example of figure \ref{fig:overview-workflow0}, the IP of the selected server (B) is retrieved from the IP Table (step \textcircled{3}).
At step \textcircled{4}, along with the reply to the connection request, the agent on server B encapsulates its load state information and its server id in the packet header.
In this \publicationtype\, the server load state is encoded inside the key option field of the GRE header~\cite{rfc2784}, which encapsulates the original IP packet\footnote{IPv6's flow id field can also be exploited to store server load information. \Albatross\ chooses the key option field of GRE header to achieve better compatibility between IPv4 and IPv6.}.
This ``passive feedback'' design differs from other LBs and reduces communication overhead with respect to periodic polling mechanisms yet keeps LBs informed before application servers reach a critical load level.
On reception of the \texttt{SYNACK} packet from the server (DSR is disabled for step \textcircled{4} with the server agent), \Albatross\ updates the load state information in the Score Table.
The packet is decapsulated and the response is forwarded back to the client (step \textcircled{6}).
The server id ($0$ in the example) is preserved in the higher bits of the TCP timestamp option.
In this way, the subsequent packets from the same flow (step \textcircled{7}) contains the server id, which helps \Albatross\ retrieve the server's IP address (step \textcircled{8}) from the IP Table and redirect immediately to the right server (step \textcircled{9}).
The server can directly answer to the client using DSR mode (step \textcircled{10}) till the end of the flow.

%% file: content/3-design.tex
\section{Design}
\label{sec:design}

\begin{figure}[t]
	\centering
	\includegraphics[width=\columnwidth]{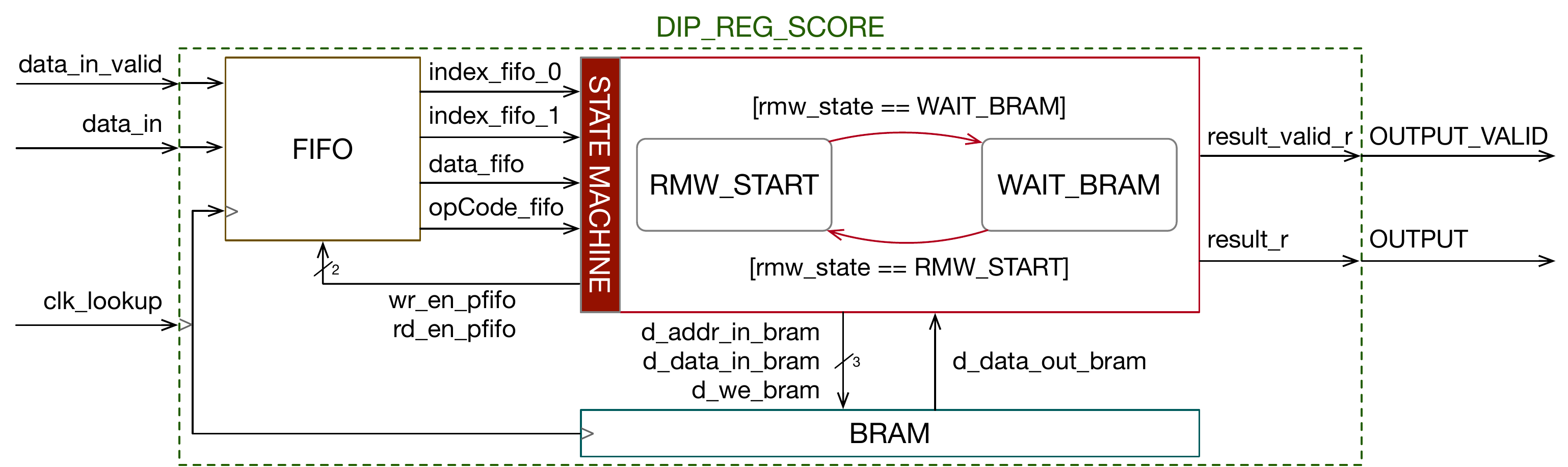}
	\caption{Schematic of \texttt{dip\_reg\_score} module.}
	\label{fig:design-module}
	\vskip -.1in
\end{figure}

The first building block of \Albatross\ is the Alias Method.
It is a probabilistic algorithm which, given initial weights, generates a table of probabilities and ``aliases'' with $\mathcal{O}(n)$ memory space complexity, where $n$ is the number of servers.
The role of the Alias Method is to distribute with higher chance the flows to servers with higher weights.
The weights are derived from servers' instant load states and are updated periodically.
The update time interval is $1$s and the choice is explained in section~\ref{sec:evaluation}.

Generating a server candidate index requires $2$ input values, \ie an entry index and a random number.
As shown in figure \ref{fig:overview-workflow0}, each entry of the Alias Table has a \textit{threshold} and an \textit{alias}.
The former determines which value is chosen, while the latter is the alternative index with respect to the entry index initially selected.
If the random number is bigger than the threshold, the output of the Alias Method is the alias, otherwise the entry index.
In the given example, four values are taken into considerations when generating $2$ server candidates.
The \texttt{idx0=0} and \texttt{idx1=2} are the two initial entry indexes of the table.
Given that the random value is smaller than the threshold $\texttt{x0}<3$, the first output is the entry index $0$.
Similarly, since $\texttt{x1}\geq9$, the second output is the alias $1$.

The $2$ values obtained from the Alias Method are then used as the ids of the $2$ server candidates. 
Their associated scores are computed with the function $g' = max(0, g - v*(\texttt{Time()} - t))$, where $g'$ is the new score, $g$ is the previous score of the server, $v$ is the ``velocity'' or the server processing speed, \texttt{Time()} is a function that returns current timestamp and $t$ is the previous timestamp.
The $3$ variables, $g$, $v$ and $t$, are saved in the Score Table.
The score $g$ is the amount of work remaining or the number of active flows on the server to execute.
The processing speed $v$ is derived from the average flow completion time (FCT) on the server side.
The timestamp $t$ corresponds to the last time the score was updated.
The time difference $\texttt{Time()} - t$ measures the elapsed time since last update.
The intuition of this function is to predict the remaining amount of tasks or queue length that a server needs to accomplish.
A higher score translates into a busier server.
The $max()$ function guarantees that the score stays non-negative. 
Once the scores of the $2$ servers are computed, the server with lower score is assigned to the flow.
In the example in figure \ref{fig:overview-workflow0}, supposing that $\texttt{Time()} = 8$ then the scores of index $0$ and $1$ are $g'_0 = max(0, 1 - 2*(8 - 5)) = 0$ and $g'_1 = max(0, 3 - 1*(8 - 7)) = 2$ respectively.
The selected server is the server with index 0, which is then mapped to server B in the IP Table.
The power-of-$2$-choices concept is applied as it has lower computational complexity than calculating the minimum yet if offers recognizable performance gains~\cite{6lb}.
For this reason, \Albatross\ better handles large-scale DCs. 


\begin{figure}[t]
	\centering
	\includegraphics[width=\columnwidth]{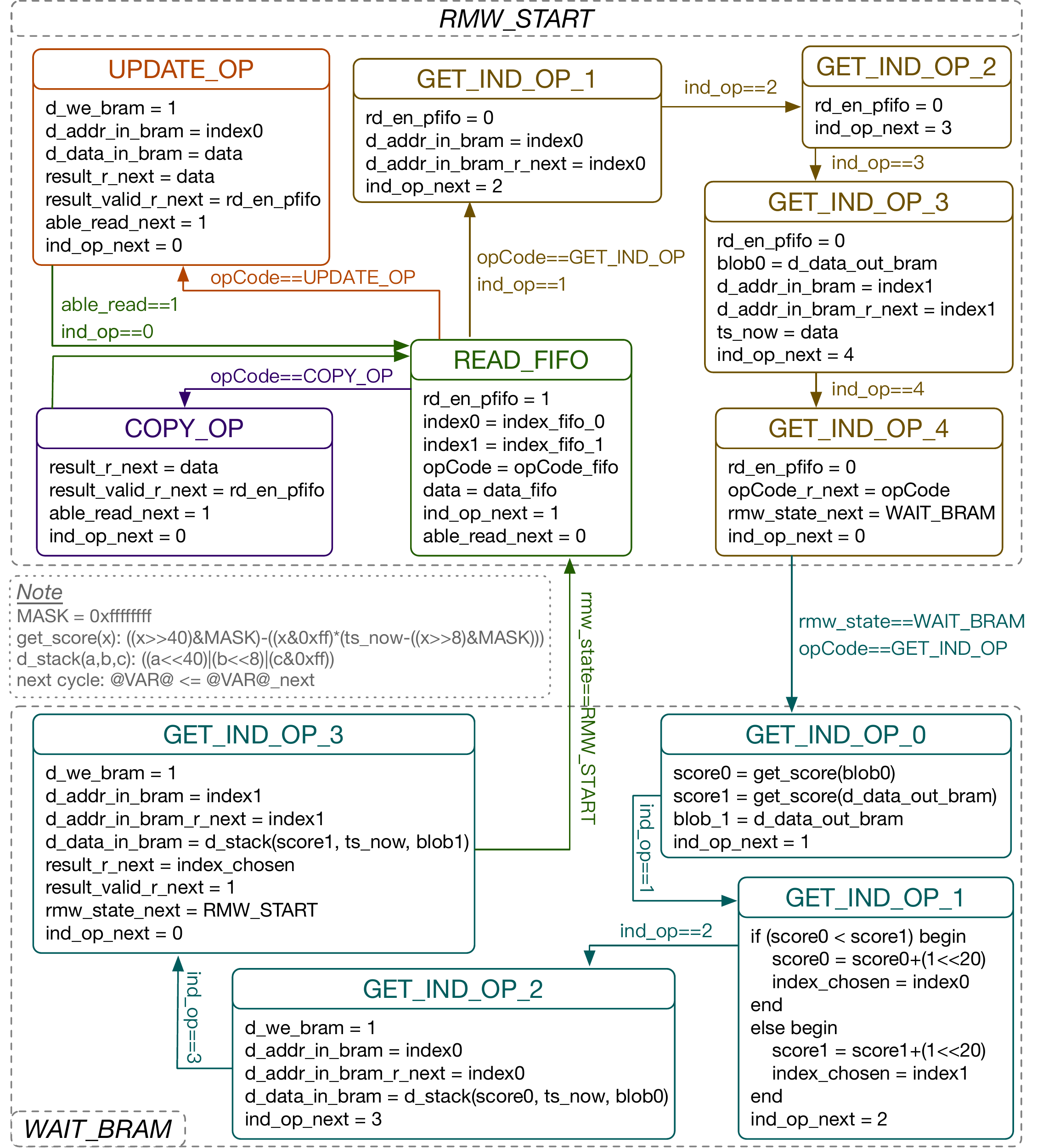}
	\caption{State machine in the \texttt{dip\_reg\_score} module.}
	\label{fig:design-state-machine}
	\vskip -.1in
\end{figure}

\begin{figure*}[t]
	\begin{center}
		\centerline{\includegraphics[width=1.8\columnwidth]{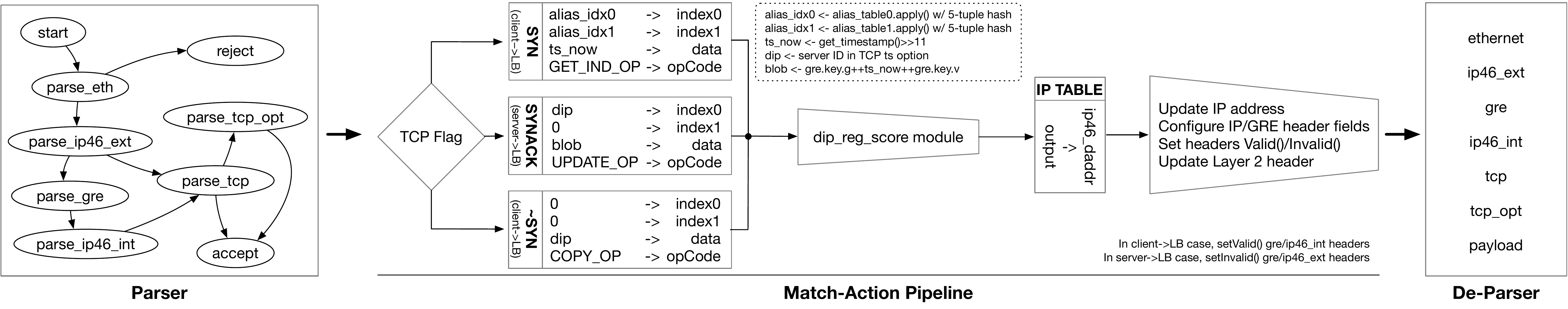}}
		\vskip -0.05in
		\caption{P4 workflow.}
		\label{fig:implement-p4-workflow}
	\end{center}
	\vskip -0.2in
\end{figure*}

The main function of \Albatross\ is implemented in a single Verilog module called \texttt{dip\_reg\_score}.
This design depends mainly on the read and write actions that should be executed on multiple indexes. 
Figure \ref{fig:design-module} shows the schematics of this module. It takes as input \texttt{data\_in\_valid}, \texttt{data\_in} and \texttt{clk\_lookup} and as output \texttt{OUTPUT\_VALID} and \texttt{OUTPUT}.
The core logic of \texttt{dip\_reg\_score} locates in the \texttt{STATE\_MACHINE} block. 
It interacts with \texttt{FIFO} and \texttt{BRAM} (Block RAM).
The former receives and stores the inputs of the block, while the latter is used to save the server load states information.
Figure \ref{fig:design-state-machine} depicts in detail the workflow of \texttt{STATE\_MACHINE}.
It consists of two states \texttt{RMW\_START} and \texttt{WAIT\_BRAM}.
Each state can be further decomposed into several states.
The initial state in \texttt{RMW\_START} is \texttt{READ\_FIFO}.
In this state the input saved in the FIFO are extracted.
Depending on the \texttt{opCode} value, $3$ operations can be executed: \texttt{UPDATE\_OP}, \texttt{COPY\_OP} and \texttt{GET\_INDEX\_OP}.
The code \texttt{UPDATE\_OP} is used to update the server load states in the Score Table given the feedback extracted from the \texttt{SYNACK} packets sent by the application servers.
The code \texttt{COPY\_OP} is a buffering operation.
It copies the data received from the input into the output.
The code \texttt{GET\_INDEX\_OP} is executed when a \texttt{SYN} packet reaches the LB.
It extracts the server load states with the $2$ given server indexes.
The reading operations require $2$ clocks for each value, which yields $4$ clocks in total for reading $2$ blobs from \texttt{BRAM}.
In the \texttt{WAIT\_BRAM} state, with the fetched blobs, the two scores are  computed and stored in the Score Table.
The server with lower score is selected and its corresponding score is incremented by $1$ unit task so that the new flow can be taken into account immediately.
The two scores are then stored back in the Score Table before \Albatross\ forwards the flow to the chosen server.

%% file: content/4-implement.tex
\section{Implementation}
\label{sec:implement}

This \publicationtype\ implements \Albatross\ using P4-NetFPGA.
P4 is a programming language in the family of Software-Defined Networking (SDN) technologies~\cite{SDN}.
It is used to program network devices' data plane and has been applied in other LBs~\cite{silkroad2017, beamer, lbas-2020}.
P4 has high flexibility. 
It can be translated into Verilog and the created module can be compiled into bitstream files using Vivado toolkit~\cite{Vivado}.

Figure \ref{fig:implement-p4-workflow} shows the workflow of \Albatross\ in P4's PISA model.
It can be split into $3$ parts: \textit{Parser}, \textit{Match-Action Pipeline} (MAP) and \textit{Deparser}.
The Parser separates different packet headers depending on the values of different fields.
Packets with unexpected packet headers are dropped while the others will be forwarded to the MAP.
In MAP, the main logic of P4 takes place using tables, which are collections of keys and values.
A possible input could be an IP address.
Using longest-prefix match, one key is matched, associated to which an action can be executed as, for instance, setting the egress port value.
Multiple tables can be applied during one packet processing.
\Albatross\ uses $2$ tables for the two server indexes obtained from the Alias Method.
The last step is the Deparser, which recollects all the header information and sends a new packet out of the egress port. 

Despite the flexibility of P4, it presents several limitations.
For instance, an external function is necessary to create memory cells to store additional information, or conduct complex operations.
Depending on the hardware targets, different languages can be used to describe these external modules.
Verilog is adopted for NetFPGA, therefore this \publicationtype\ implement the external module \texttt{dip\_reg\_score} in Verilog.
The core logic of this module is designed for \texttt{SYN} and \texttt{SYNACK} packets.
It serves as a buffer for other packet types.
The other external modules of \Albatross\ implemented in Verilog are the IP table, current timestamp calculation and server id extraction from the TCP timestamp option.
The IP table is a fundamental component used to redirect packets coming from the client.
Timestamp calculation takes place when a score computation or update happens.
Server id extraction is used for any packet traversing the LB with the presence of TCP timestamp option, except for \texttt{SYN} packets because in this case the LB has not yet assigned any server to the flow.

In this \publicationtype\, these external modules are configured as follows.
The target number of servers is defined as $16$, which yields $\mathcal{O}(16)$ memory space complexity with the $3$ tables\footnote{$16$ is small yet can be updated at ease.}.
The FIFO memory inside the \texttt{dip\_reg\_score} module has a queue length of $64$.
As a small-scale prototype implementation, the Vivado simulations have been applied only to $1$ of the $4$ possible Ethernet interfaces.
Another assumption of this \publicationtype\ regards the server id encapsulation in the TCP timestamp option.
To be able to encode up to $16$ servers, the server id takes $4$ higher bits in the total $32$-bit timestamp value\footnote{The assumed maximum number of bits used for encoding server id is $8$, \ie $256$ servers in total, which is sufficient for modern DCs~\cite{facebook-dc-architecture}. Any change in the timestamp value that modifies more than 24 bits is ignored.}.
To simplify the P4 code, the only option considered in the TCP header is the timestamp option.


%% file: content/5-evaluation.tex
\section{Evaluation}
\label{sec:evaluation}

This section evaluates \Albatross\ from $3$ perspectives, (i) acceptance rate of covert channel existence in packet headers, (ii) performance gain in terms of load balancing fairness and QoS, and (iii) throughput and additional processing latency using P4-NetFPGA implementation.

\subsection{Covert Channel Acceptance}
\label{sec:evaluation-covert}

To understand how the Internet would react to the presence of timestamp option in the TCP header, requests are sent from Paris to over $60$k distinct random IP addresses.
The results obtained are the following:

\begin{itemize}
    \item \texttt{NO CONNECTION} = $45019$
    \item \texttt{SUCCESS} = $12876$
    \item \texttt{FAILURE} = $5787$
    \item \texttt{TOTAL} = $63682$
\end{itemize}

\begin{figure*}[t]
	\vskip -0.05in
	\centering
	\begin{subfigure}{\columnwidth}
		\centering
		\includegraphics[width=.95\columnwidth]{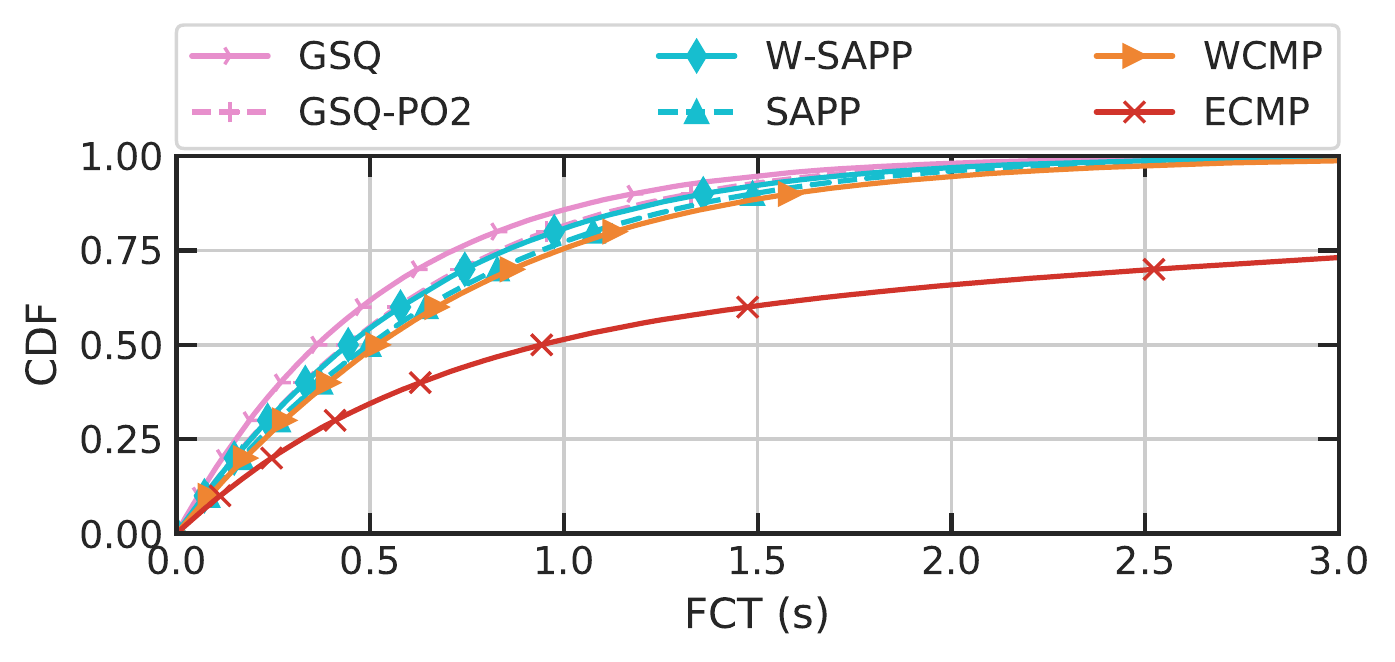}
		\vskip -0.05in
		\caption{$64.5\%$ expected resource utilization.}
		\label{fig:evaluation-fairness-645}
	\end{subfigure}
	\hspace{.05in}
	\begin{subfigure}{\columnwidth}
		\centering
		\includegraphics[width=.95\columnwidth]{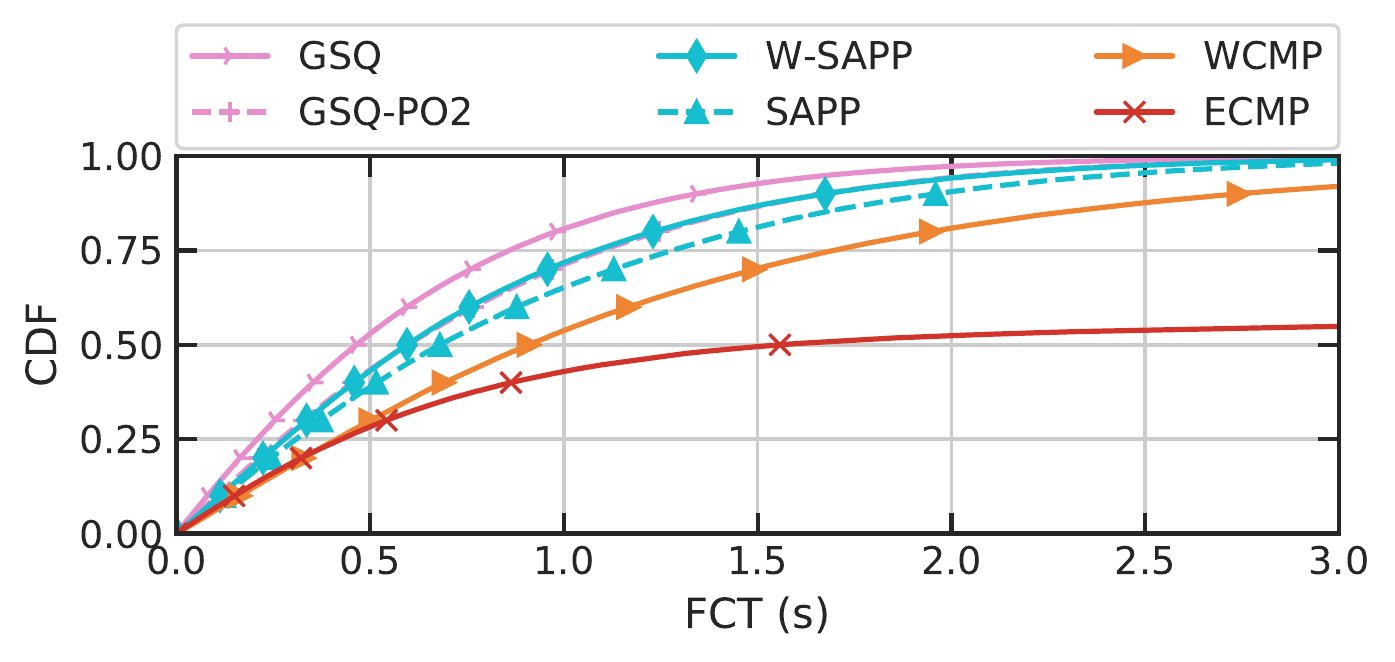}
		\vskip -0.1in
		\caption{$84.5\%$ expected resource utilization.}
		\label{fig:evaluation-fairness-845}
	\end{subfigure}
	\vskip .1in
	\begin{subfigure}{\columnwidth}
		\centering
		\includegraphics[width=.95\columnwidth]{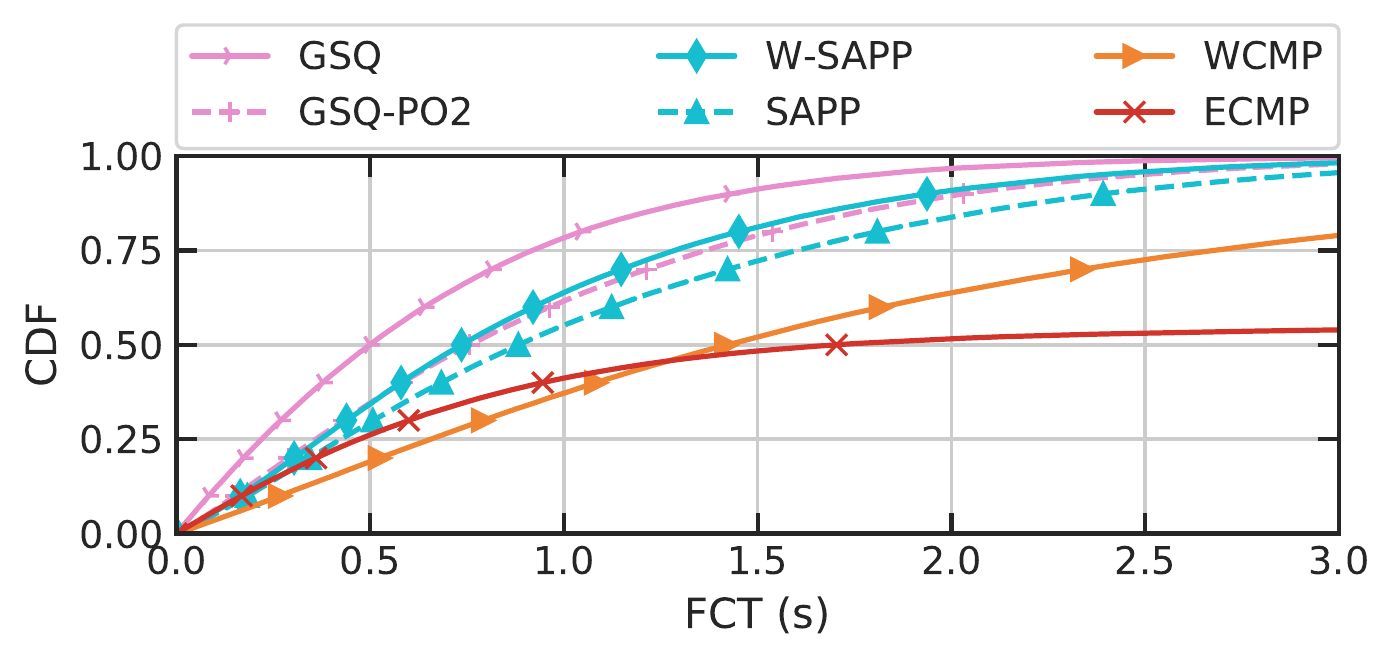}
		\vskip -0.1in
		\caption{$92.5\%$ expected resource utilization.}
		\label{fig:evaluation-fairness-925}
	\end{subfigure}
	\hspace{.05in}
	\begin{subfigure}{\columnwidth}
		\centering
		\includegraphics[width=.95\columnwidth]{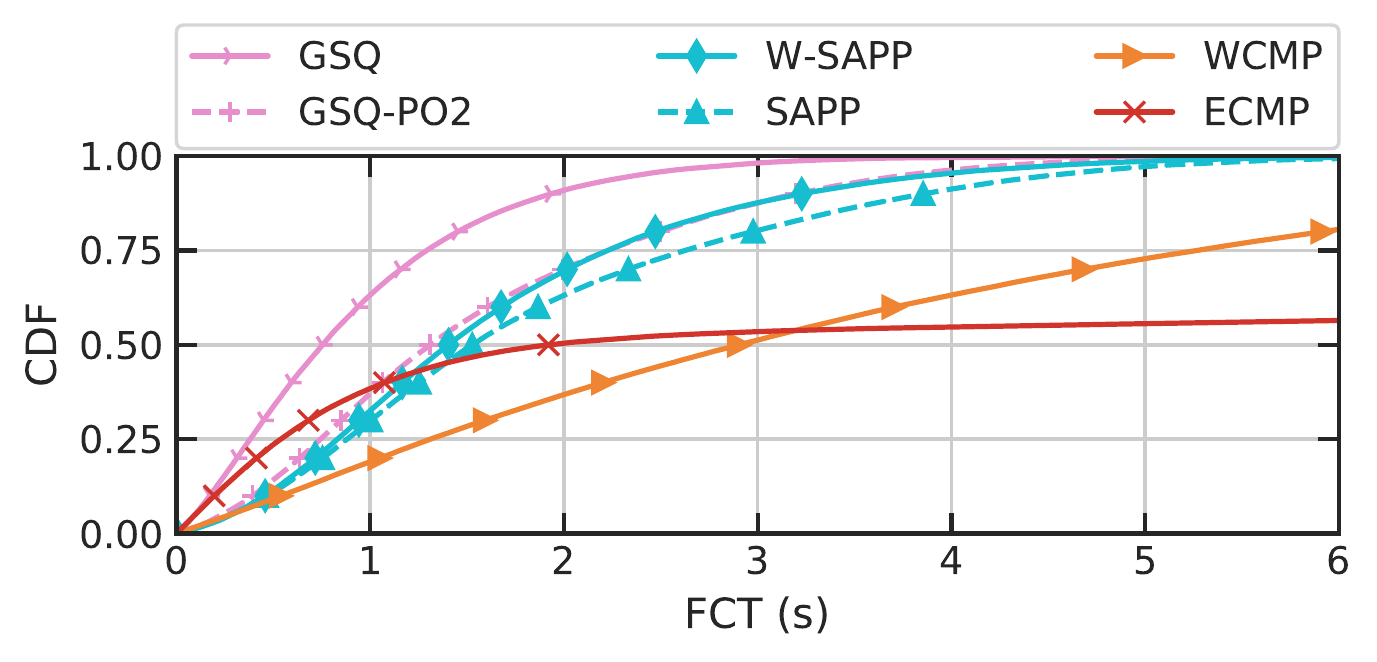}
		\vskip -0.1in
		\caption{$100\%$ expected resource utilization.}
		\label{fig:evaluation-fairness-1000}
	\end{subfigure}
	\caption{CDF of FCT of different LB designs at various traffic rate.}
	\label{fig:evaluation-fairness}
	\vspace{-.2in}
\end{figure*}

The code \texttt{NO CONNECTION} is the number of connections which have not received any response regardless of the presence of the timestamp option.
The code \texttt{SUCCESS} is the number of connections that have answered to a packet with the timestamp option.
The code \texttt{FAILURE} is the number of connections that have not answered to a packet with the timestamp option but answered to packets without timestamp option.
Pruning the first case where the IP addresses can not associated to any device or service is not available and analyzing only \texttt{SUCCESS} and \texttt{FAILURE} cases gives an acceptance rate of $68.99\%$.
This experiment does not study the different geographic locations of the clients and servers or other factors yet it validates that the stateless design of \Albatross\ works for most end hosts.
It is also in accordance with the high acceptance rate (over $86\%$) obtained by experimenting on a larger scale of testbed in~\cite{honda2011still}.

\subsection{Load Balancing Fairness}
\label{sec:evaluation-fairness}

A simulator is built with $2$ LBs and $64$ application servers with different processing capacities\footnote{Half of the application servers have $2$ times higher processing capacities than the other half.} to study the load balancing performance in terms of workload distribution fairness. 
$3$ episodes of $50$k flows that last $500$ms on average are simulated as Poisson traffic at different variances.
The traffic rates are normalized by the total server cluster processing capacities.
Figure \ref{fig:evaluation-fairness} depicts the cumulative distribution function (CDF) of FCT.
Different LB designs are compared.
Global shotest queue (GSQ), as the name suggests, chooses the application server with the shortest queue.
It is an oracle solution that can be achieved assuming that the LBs are aware of the actual queue lengths on each server and no computational overhead is incurred when computing the minimum queue length.
It represents the best performance a LB can achieve with perfect server load information.
GSQ-PO2 applies power-of-2-choices over GSQ.
Similar to GSQ, the LBs are assumed to be aware of the exact instant queue lengths on each server.
Unlike GSQ, GSQ-PO2 selects $2$ random server candidates and then picks the one with a shorter queue.
It represents the theoretical best performance \Albatross\ can achieve.
ECMP randomly selects the application servers and is the most widely employed load balancing mechanism.
WCMP selects the application servers according to its statically configured weights which are proportional to the server processing capacities.
W-SAPP denotes the implementation of \Albatross.
SAPP corresponds to a simplified version of \Albatross, where the $2$ server candidates are chosen using a uniform distribution instead of using weighted sampling with the Alias method.
The main difference between SAPP and W-SAPP is the probabilistic method that W-SAPP applies to obtain the $2$ choices.
The weights, which are used later as probabilities, are defined using the relative processing capacities of application servers. 
As depicted in figure~\ref{fig:evaluation-fairness}, the performances of both SAPP and W-SAPP are similar to GSQ-PO2, which is considered as the method that takes the perfect choice.
Improvements can be observed for W-SAPP over SAPP, which shows the added value of using the Alias Method in \Albatross.

\begin{table}[t]
    \centering
     \begin{tabular}{|c |c |c |c |c |c |c |} 
     \hline
     Rates & GSQ & GSQ2 & W-SAPP & SAPP & WCMP & ECMP\\
     \hline
        $64.5\%$ & $0.91$ & $0.92$ & $0.89$ & $0.91$ & $0.71$ & $0.50$\\
        \hline
        $76.5\%$ & $0.92$ & $0.94$ & $0.91$ & $0.93$ & $0.71$ & $0.54$\\ 
        \hline
        $84.5\%$ & $0.93$ & $0.95$ & $0.93$ & $0.94$ & $0.69$ & $0.55$\\ 
        \hline
        $92.5\%$ & $0.93$ & $0.96$ & $0.94$ & $0.96$ & $0.65$ & $0.56$\\ 
        \hline
        $100\%$ & $0.95$ & $0.98$ & $0.97$ & $0.97$ & $0.71$ & $0.57$\\ 
        \hline
    \end{tabular}
    \vskip .1in
    \caption{Jain's Fairness indexes of different LBs at different traffic rates.}
    \label{tab2}
    \vskip -.3in
\end{table}

Another metric to evaluate load balancing fairness is the Jain's fairness index~\cite{Jain_index}, which computes the fairness of workload distribution.
Considering $n$ servers each one with a particular amount of flows $x_i$, the fairness index is computed as $\left(\frac{\sum_{i=1}^{n} x_i}{n}\right)^2 \cdot \left(\frac{\sum_{i=1}^{n} x_i^2}{n^2}\right)^{-1}$.
The maximum and minimum values that the index can reach are respectively $1$ and $\frac{1}{n}$.
If the index reaches value $1$, it means that the load has been fairly distributed.
The worst case is when the index is equal to $\frac{1}{n}$ which proves that only one server has taken all the flows. 

Using the same configuration as in previous simulations, the fairness indexes of different LB designs are computed.
As shown in table \ref{tab2}, ECMP and WCMP have the worst performance.
Random choices do not guarantee a fair distribution of flows.
On the other hand, GSQ and GSQ-PO2 get the best fairness.
They have perfect knowledge of the server states.
W-SAPP and SAPP achieve similar performance to GSQ and GSQ-PO2.
Although SAPP achieves a better fairness, it has to be taken into consideration that the probabilities that W-SAPP uses to choose server candidates are statically configured proportional to server processing capacities.
The Alias Method in \Albatross\, however, uses dynamic weights to select a subset of candidates.

Another important parameter to analyze is the update time intervals of the Alias Table.
If the update time interval is too high, the LB choice would not reflect the real-time load states of the application servers.
For this reason, different update time intervals are simulated.

\begin{figure}[t]
	\vskip -0.05in
	\centering
	\begin{subfigure}{\columnwidth}
		\centering
		\includegraphics[width=.95\columnwidth]{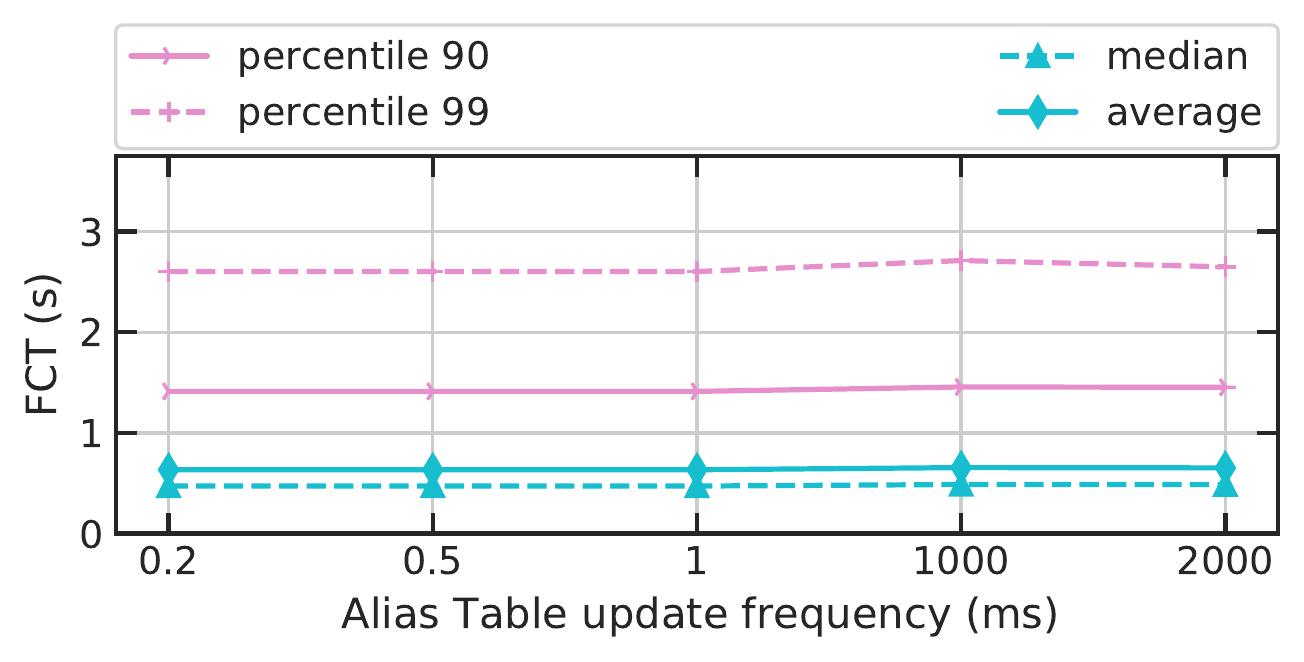}
		\vskip -0.1in
		\caption{$64.5\%$ expected resource utilization.}
		\label{fig:evaluation-frequency-645}
	\end{subfigure}
	\vskip 0.05in
	\begin{subfigure}{\columnwidth}
		\centering
		\includegraphics[width=.95\columnwidth]{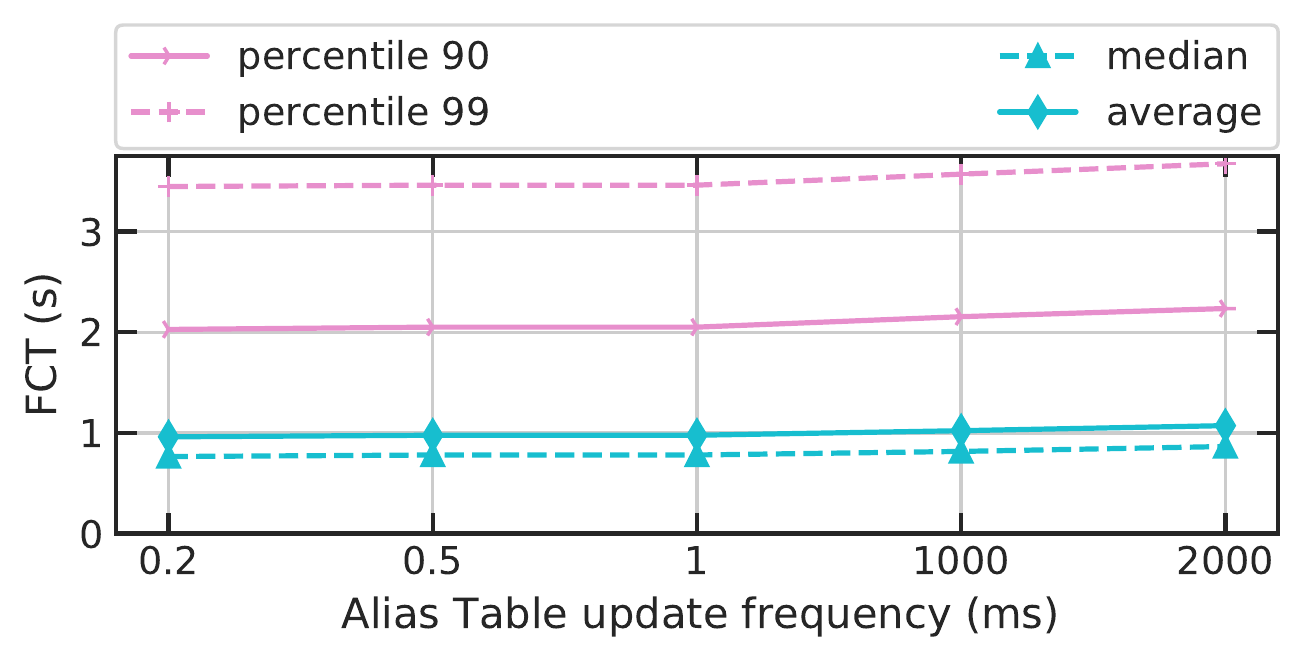}
		\vskip -0.1in
		\caption{$92.5\%$ expected resource utilization.}
		\label{fig:evaluation-frequency-925}
	\end{subfigure}
	\caption{FCT of different Alias Table update interval LB designs at various traffic rate.}
	\label{fig:evaluation-frequency}
\end{figure}

The results of the simulations are depicted in figure \ref{fig:evaluation-frequency}.
$4$ values are plotted at $2$ different traffic rates: percentile $90$, percentile $99$, median and average of FCT. 
Five different time intervals of Alias Table updates are used: $0.2$ ms, $0.5$ ms, $1$ ms, $1$ s and $2$ s.
The plots show a slight improvement of FCT when the update time interval is lower (higher update frequency).
This difference is too small to justify shorter time interval update.
The LBs are not significantly influenced by a real-time update of the weights.
However, these are simulation results.
Physical experiments are necessary to further justify this assumption.

\subsection{P4-NetFPGA Implementation Performance}
\label{sec:evaluation-p4}

\begin{figure}[t]
	\begin{center}
		\vskip -0.1in
		\centerline{\includegraphics[width=0.95\columnwidth]{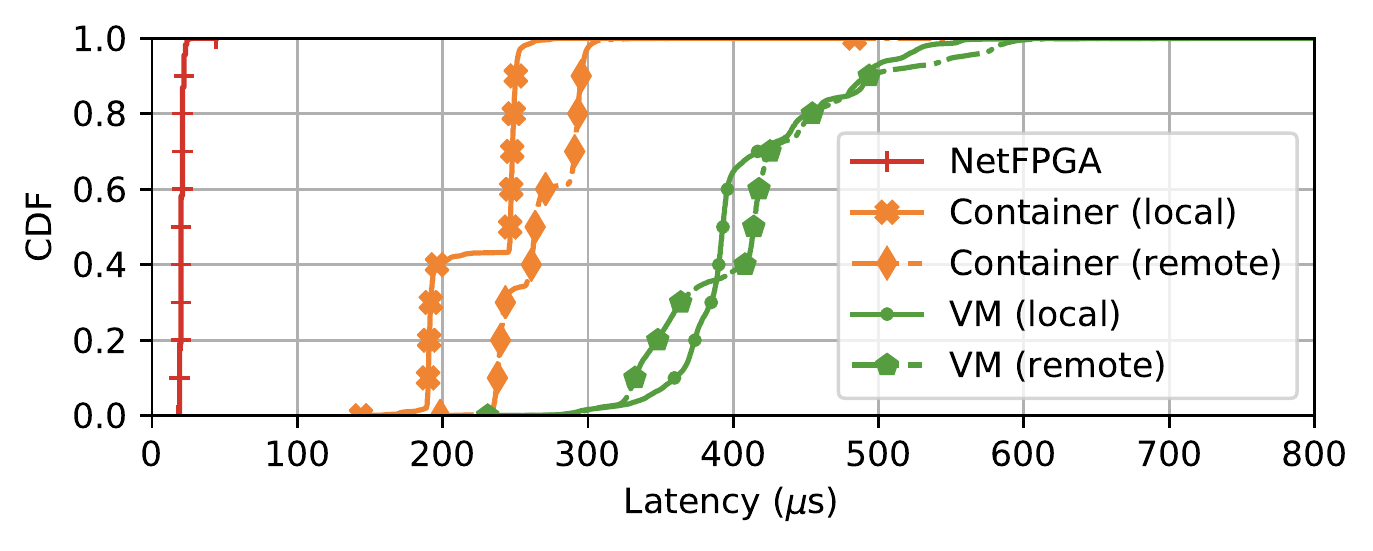}}
		\vskip -0.1in
		\caption{Latency CDF of different implementation.}
		\vskip -0.2in
		\label{fig:evaluation-p4-latency}
	\end{center}
\end{figure}

The performance of NetFPGA is compared with respect to software solutions. 
On the NetFPGA, the \texttt{reference\_NIC} bitstream file is loaded, making the NetFPGA behaves as a NIC.
This program returns on the PCI the packets that the NetFPGA receives from the Ethernet interface.
In the same way, a packet sent to the PCI is then sent from the Ethernet interface.
This allows to find at which timestamp the packet sent through the PCI traverses the Ethernet interface and reaches the host machine. 
Figure \ref{fig:evaluation-p4-latency} shows the CDF of the latency.
The Round-Trip Time (RTT) of ping packets of other software solutions are also depicted. 
In particular, four cases are considered: when there are two containers (docker) or two VMs (KVM) on the same machine or on different machines.
The performance of the NetFPGA largely outperforms the software solutions, which makes NetFPGA a preferred solution in terms of performance.

\begin{figure}[t]
	\begin{center}
		\vskip -0.1in
		\centerline{\includegraphics[width=0.95\columnwidth]{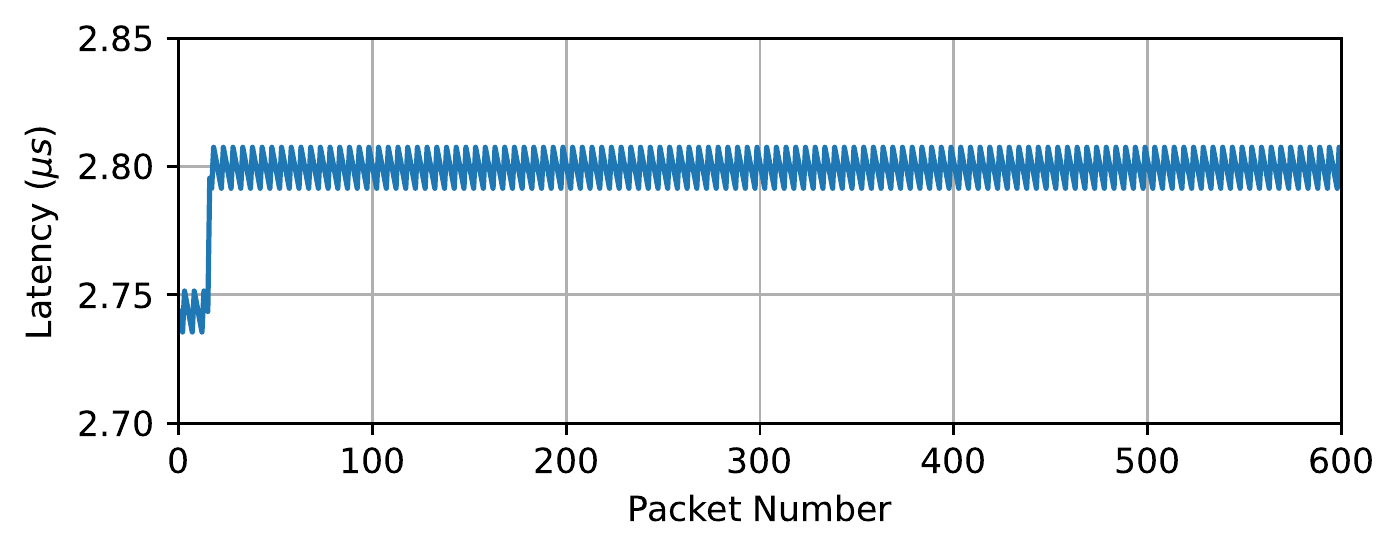}}
		\vskip -0.1in
		\caption{Delay in packet departure with respect to the number of packets sent.}
		\vskip -0.2in
		\label{fig:evaluation-throughput}
	\end{center}
\end{figure}

To demonstrate the performance of the Verilog module \texttt{dip\_reg\_score}, Vivado behavioural simulations is conducted.
A burst of $600$ packets is sent to the NetFPGA board.
The delay between packet arrival and departure is shown in figure \ref{fig:evaluation-throughput}.

The difference between the first $16$ packets and the remaining packets is due to the nature of the packets sent.
The first batch of $16$ packets traverse a shorter path as they upload the IP addresses in the IP table.
The remaining packets are TCP \texttt{SYN} packets.
They traverse the longest path of the \texttt{dip\_reg\_score} module, in which the destination application server of the flow is chosen. 
The delay is almost linear to the number of cycles required for packet processing and stays constant.
The sinusoidal shape is because of the jitter, which is low.
This plot shows the high performance of the designed module and its efficiency in executing the proposed LB algorithm.

%% file: content/6-conclusion.tex
\section{Conclusion}
\label{sec:conclusion}

This paper proposes \Albatross\, a stateless, load-aware, hardware load balancer in DCs, which (i) fairly distributes connections' requests, (ii) guarantees PCC, and (iii) minimizes additional latency due to its presence.
The design choices of \Albatross\ makes it suitable for implementation on programmable hardware.
Simulation results show \Albatross\ improves load balancing fairness and helps achieve better quality of service than other LB mechanisms.
Evaluations of throughput and processing latency demonstrate the advantage of hardware implementations.